\documentclass[floatfix,superscriptaddress,showpacs,amssymb,10pt,aps,prd,reprint,longbibliography]{revtex4-1}

\usepackage{graphicx,epsfig,amssymb} 
\usepackage{amsmath,amsfonts, times}
\usepackage{bm} 

\usepackage[linktocpage,colorlinks]{hyperref}
\usepackage[caption=false]{subfig}
\usepackage[usenames]{color}     
\usepackage{natbib}
\usepackage{soul}
\usepackage[utf8x]{inputenc}
\usepackage{float}

\definecolor{coolblack}{rgb}{0.0, 0.18, 0.39}
\definecolor{darkred}{rgb}{0.5,0,0}
\definecolor{darkgreen}{rgb}{0,0.5,0}
\definecolor{darkblue}{rgb}{0,0,0.5}
\definecolor{lapislazuli}{rgb}{0.15, 0.38, 0.61}
\definecolor{venetianred}{rgb}{0.78, 0.03, 0.08}
\definecolor{bleudefrance}{rgb}{0.19, 0.55, 0.91}
\definecolor{dogwoodrose}{rgb}{0.84, 0.09, 0.41}
\hypersetup{colorlinks=true, citecolor=darkgreen, linkcolor=darkblue, 
urlcolor = blue}

\def\be{\begin{equation}}
\def\ee{\end{equation}}

\renewcommand{\vec}[1]{\boldsymbol{#1}}

\newcommand{\bea}{\begin{eqnarray}}
\newcommand{\eea}{\end{eqnarray}}
\newcommand{\ben}{\begin{enumerate}}
\newcommand{\een}{\end{enumerate}}
\newcommand{\bi}{\begin{itemize}}
\newcommand{\ei}{\end{itemize}}

\newcommand{\thet}{{{\theta}}}

\def\ga{\mathrel{\raise.3ex\hbox{$>$\kern-.75em\lower1ex\hbox{$\sim$}}}}
\def\la{\mathrel{\raise.3ex\hbox{$<$\kern-.75em\lower1ex\hbox{$\sim$}}}}

\def\l{\left}
\def\r{\right}
\def\be{\begin{equation}}
\def\ee{\end{equation}}
\def\b{\begin{equation}}
\def\e{\end{equation}}

\renewcommand{\vec}[1]{\boldsymbol{#1}}

\def\I_M{{I_{\scriptscriptstyle M\times M}}}

\def\be{\begin{equation}}
\def\ee{\end{equation}}
\def\bea{\begin{eqnarray}}
\def\eea{\end{eqnarray}}
\newcommand{\beq}{\begin{eqnarray}}
\newcommand{\eeq}{\end{eqnarray}}

\renewcommand{\d}{\rm{d}}
\newcommand{\beqal}{\begin{eqnarray}\label}
\newcommand{\beqa}{\begin{eqnarray}}
\newcommand{\eeqa}{\end{eqnarray}}

\begin{document}
\title{\large Massive and charged scalar field in Kerr-Newman spacetime: Absorption and superradiance}


\author{Carolina L. Benone}
\email{lben.carol@gmail.com}
\affiliation{Campus Universitário Salinópolis, Universidade Federal do Pará, 68721-000, Salinópolis, Pará, Brazil.}
\affiliation{Faculdade de F\'{\i}sica, Universidade 
Federal do Par\'a, 66075-110, Bel\'em, Par\'a, Brazil.}

\author{Lu\'is C. B. Crispino}
\email{crispino@ufpa.br}
\affiliation{Faculdade de F\'{\i}sica, Universidade 
Federal do Par\'a, 66075-110, Bel\'em, Par\'a, Brazil.}

\begin{abstract}
We consider the propagation of a generic scalar field around a rotating and charged black
hole. Using the partial wave method, we find, numerically, the total and partial absorption cross sections for
different incidence angles. We investigate the low- and high-frequency limits, finding semi-analytical approximations
for the absorption cross section, which we compare with our numerical results. Finally, we consider the superradiant regime, showing that, for charged fields, planar waves can be superradiantly scattered.

\end{abstract}

\date{\today}

\maketitle

\section{Introduction}\label{sec:int}

General relativity (GR) is the most accepted theory of gravity, passing every test so far \cite{Will:2014kxa}. However, most tests were performed in the weak field regime. This scenario changed with the detections of gravitational waves by the LIGO/VIRGO collaborations \cite{GWs,Abbott:2016nmj,Abbott:2017vtc,Abbott:2017gyy,Abbott:2017oio,TheLIGOScientific:2017qsa}, what confirmed a prediction of GR for the strong field regime. Such detections allow us to access the dynamics of dramatic phenomena, such as the merger of two black holes. 


Besides the detection of black holes through gravitational waves, the shadows of these objects are expected to be probed in the near future with the Event Horizon Telescope (EHT) \cite{Psaltis:2018xkc}. This could help to test the no-hair conjecture, by checking deviations from the Kerr geometry \cite{Johannsen:2010bi}. Furthermore, with the information from the vicinity of the event horizon, one could also access horizon-scale quantum fluctuations \cite{Giddings:2016btb}.

Scattering and absorption processes play important roles in the study of black holes. Shadows of black holes, in particular,  are obtained when we consider the scattering of light by a black hole. For fields around rotating black holes, one finds that low-frequency modes are subjected to superradiant scattering \cite{PhysRevLett.28.994}, what decreases the energy of the black hole. This effect is not exclusive of black holes, being present also in other contexts \cite{Dicke:1954zz, zeldovich1, zeldovich2, Oliveira:2010zzb, Dolan:2011zza}. Although it has been predicted in the 70's, rotational superradiance was only recently verified in the laboratory \cite{Torres:2016iee}.

Superradiance is also present in the context of static black holes, when we consider, for instance, charged fields around Reissner-Nordström black holes \cite{DiMenza:2014vpa,Benone:2015bst}. This phenomena is much more efficient for the charged case than for the rotating case. If we consider, for instance, the scalar field case, the maximum amplification for the rotating case is $0.4\%$, while for the charged case it reaches $40\%$ for a massless field with charge $q=1$ \cite{brito2015superradiance}.

Superradiance is associated to a negative absorption cross section. For a rotating black hole, the scalar absorption cross section can be negative for corotating spherical waves, remaining positive for planar waves \cite{Caio:2013}. Nevertheless, for a static charged black hole, due to the contribution of the Lorentz force, the absorption cross section can be negative for planar waves \cite{Benone:2015bst}. 

Various works on scattering and absorption of fields by static black holes can be found in the literature, for uncharged \cite{matzner1968scattering,Sanchez:1977si,Jung:2004yh,Dolan:2006vj,Doran:2005vm} and charged black holes \cite{Crispino:2008zz,Benone:2014qaa,PhysRevD.95.044035,Benone:2015bst}. 
For stationary black holes there are works considering spin 0 \cite{Macedo:2013afa,Leite:2017hkm,Benone:2018}, spin 1 \cite{Leite:2017zyb, Leite:2018mon} and spin 2 fields \cite{Dolan:2008kf,Leite:2017zyb}. However, regarding the Kerr-Newman spacetime there is still a wide gap in the investigations of absorption.

The study of charged black holes is sometimes left aside, due to the consideration that astrophysical black holes are not expected to have significant charge \cite{Gibbons:1975kk}. However, studies of minicharged dark matter suggest the possibility of charged black holes \cite{Cardoso:2016olt}. In any case, from a theoretical point of view, the study of charged solutions gives a more complete picture of the physics of black holes.

We consider the absorption of a massive and charged scalar field by a Kerr-Newman black hole. Due to the complexity of the corresponding equations, the solutions are obtained using a numerical approach, but for the low- and high-frequency regimes we obtain semi-analytical approximations. We pay special attention to the role of the charge in the superradiance.

The remaining of this work is organized as follows. In Sec. \ref{sec:scalarfield} we present the relevant equations for computing the absorption of a massive charged field around a charged and rotating black hole. In Sec. \ref{sec:hfl} we consider the equations of motion of a massive and charged particle in a Kerr-Newman spacetime, finding a semi-analytical approximation for the absorption cross section. In Sec. \ref{sec:low} we consider the low-frequency limit of the field equations, obtaining a generalization for the (well-known) corresponding limit of the absorption cross section. In Sec. \ref{sec:res} we present our numerical results for the partial and total absorption cross sections, which we compare with the semi-analytical results obtained in previous sections. In Sec. \ref{sec:remarks} we conclude with our final remarks. We use natural units, so that $c=G=\hbar=1.$

\section{Scalar absorption in Kerr-Newman spacetimes}\label{sec:scalarfield}
The Kerr-Newman spacetime can be described by the line element
\bea
& d s^2&= \l(1-\frac{2Mr-Q^2}{\rho^2}\r)d t^2-\frac{\rho^2}{\Delta}d r^2-\rho^2 d\theta^2 \nonumber\\
&+&\frac{4Mar\sin^2\theta-2aQ^2\sin^2\theta}{\rho^2}d t \d\phi-\frac{\xi\sin^2\theta}{\rho^2}d\phi^2,\label{eq:linelement}
\eea
where ~$\rho^2= r^2+a^2\cos^2\theta$, $\Delta= r^2-2Mr+a^2+Q^2$, and $\xi= (r^2+a^2)^2-\Delta a^2\sin^2\theta$. For $M\geq \sqrt{a^2+Q^2}$, this spacetime is associated to a black hole that possesses mass $M$, charge $Q$ and angular momentum $J=aM$. This solution generally presents two horizons, $r_\pm=M\pm\sqrt{M^2-(a^2+Q^2)}$, which coincide for $M^2=a^2+Q^2$.

We are interested in the absorption of a massive and charged scalar field, so that we have to solve the following Klein-Gordon equation 
\begin{equation}
(\nabla^\alpha+iq A^\alpha)(\nabla_\alpha+iq A_\alpha)\Psi+\mu^2 \Psi =0.
\label{eq:kgeq}
\end{equation}
Equation (\ref{eq:kgeq}) dictates the dynamics of a scalar field with charge $q$ and mass $\mu$ ($<\omega$), subjected to the electromagnetic vector potential
\b
A_\alpha = \frac{rQ}{\rho^2}(1,0,0,-a\sin^2{\theta}).
\label{epot}
\e

In order to solve Eq. (\ref{eq:kgeq}), we perform the separation of variables given by
\be
\Psi_\omega=\sum_{l=0}^{+\infty}\sum_{m=-l}^{+l}\frac{U_{\omega lm}(r)}{\sqrt{r^2+a^2}}S_{\omega lm}(\theta)e^{im\phi-i\omega t},
\label{eq:fieldecomposition}
\ee
where $S_{\omega l m}$ are the spheroidal harmonics, which obey the equation
\bea
&\l(\frac{d^2}{d\thet^2}+\cot\thet\frac{d}{d\theta}\r)S_{\omega lm}\nonumber\\
&+\l[\lambda_{lm}+a^2(\omega^2-\mu^2)\cos^2\theta-\frac{m^2}{\sin^2\theta}\r]S_{\omega lm}=0,\label{eq:spheroidaleq}
\eea
with~$\lambda_{lm}$ being the eigenvalues of the spheroidal harmonics, which are normalized according to
\be
\int d\theta\,\sin\theta\,\l|S_{\omega lm}(\theta)\r|^2=\frac{1}{2\pi}.
\ee

We define the tortoise coordinate as 
\be
r_{\star}\equiv\int d r\,\l(\frac{r^2+a^2}{\Delta}\r)
\label{eq:tortoisecoord}
\ee
and use it to rewrite the radial equation as
\be
\l(\frac{d^2}{d r_\star^2}+V_{\omega lm}\r)U_{\omega lm}(r_\star)=0,
\label{eq:radialeq}
\ee
with~$V_{\omega lm}$ given by
\begin{widetext}
\be
V_{\omega lm}(r)= \frac{H^2+ [2ma\omega-\mu^2 (r^2+a^2) - \lambda_{l m}-a^2(\omega^2-\mu^2)]\Delta}{(r^2+a^2)^2}-\frac{\Delta[\Delta+2r(r-M)]}{(r^2+a^2)^3}+ \frac{3r^2\Delta^2}{(r^2+a^2)^4},
\label{eq:effpot}
\ee
\end{widetext}
where
\be
H\equiv(r^2+a^2)\omega - am - qQr.
\ee

The asymptotic limits of the solution of Eq. (\ref{eq:radialeq}), associated to an incident wave from spatial infinity, are given by
\be
U_{\omega lm}(r_\star)\sim\left\{
\begin{array}{c l}
	{\mathcal{I}_{\omega lm}}U_I+{{\cal R}_{\omega lm}} U_I^* & (r_\star/M\rightarrow +\infty),\\
	{\mathcal{T}_{\omega lm}} U_T & (r_\star/M\rightarrow -\infty),
\end{array}\right.
\label{inmodes}
\ee  
in which
\be
U_I = e^{-i \omega v r_\star}r^\tau\sum_{j=0}^N \frac{h_j}{r^j}
\label{eq:inc}
\ee
is associated to an incident wave, with
\be
\tau \equiv \frac{i (M\mu^2- qQ\omega)}{\omega v} 
\ee
and
\be
v\equiv \sqrt{1-\frac{\mu^2}{\omega^2}}.
\ee
The transmitted wave is given by
\be
U_T = e^{-i \left({\omega-\omega_c}\right)r_\star}\sum_{j=0}^N g_j (r-r_+)^j,
\label{eq:trw}
\ee
where
\be
\omega_c \equiv m \frac{a}{r_+^2+a^2} + \frac{q Q r_+}{r_+^2+a^2}
\ee
is the critical frequency.

For $\omega>\omega_c$, $U_T$ corresponds to a wave going into the black hole horizon. However, for $\omega<\omega_c$ the exponential of $U_T$ changes its overall sign, corresponding to an outgoing wave. In the latter case we have superradiance, what will be investigated in Sec. \ref{sec:res}.

Using the partial wave method one can show that the total absorption cross section can be written as
\b
\sigma=\sum_{l=0}^\infty\sum_{m=-l}^l \sigma_{l,m},
\label{abs}
\e
where $\sigma_{l,m}$ is the partial absorption cross section, given by
\b
\sigma_{l,m}=\frac{4\pi^2}{\omega^2 v^2}|S_{\omega l m}(\gamma)|^2 \left[1-\left|\frac{{\mathcal{R}_{\omega lm}}}{{\mathcal{I}_{\omega lm}}}\right|^2\right],
\label{eq:abs}
\e
with $\gamma$ being the incidence angle. We can split $\sigma$ into corotating modes, given by
\b
\sigma^{\text{co}}=\sum_{l=1}^\infty\sum_{m=1}^l \sigma_{l, m},
\label{coro}
\e
counterrotating modes, given by
\b
\sigma^{\text{counter}}=\sum_{l=1}^\infty\sum_{m=1}^l \sigma_{l, -m},
\label{counter}
\e
and axially-symmetric modes, 
\b
\sigma^{\text{axial}}=\sum_{l=0}^\infty \sigma_{l, 0}.
\label{axial}
\e
Such separation is useful for understanding the behavior of off-axis incidence.

\section{High-frequency limit}\label{sec:hfl}

In the high-frequency limit we can find an approximation for the absorption cross section by analyzing the path of a massive and charged particle subjected to the electromagnetic vector potential (\ref{epot}). For a wavefront impinging upon the black hole, making an angle of $\gamma$ with the rotation axis, the high-frequency absorption cross section is given by \cite{Macedo:2013afa}
\be
\sigma_{\text{hf}}=\int_{-\pi}^\pi b_c(\chi,\gamma) d\chi,
\ee
where $\chi$ is an angle defined in the plane of the wavefront and $b_c$ is the critical impact parameter. 

In order to find the critical impact parameter, we consider an \textit{impact vector} given by 
\be
\vec{b}=\l(b\cos\gamma \cos\chi,b\sin\chi,-b\sin\gamma\sin\chi\r).
\label{biv}
\ee
Considering Eq. (\ref{biv}) we find, for the radial part of the equations of motion,
\be
R(r)=P^2 - \Delta[(1-v^2)r^2+(\mathcal{L}_z-a)^2+\mathcal{C}], 
\ee
where
\be
P\equiv(r^2+a^2)-a \mathcal{L}_z - \frac{q Q r}{E},
\ee
with
\be
\mathcal{L}_z\equiv L_z E^{-1}=b v \sin{\chi}\sin{\gamma}, 
\ee
and 
\be
\mathcal{C}\equiv\frac{C}{E^{2}}=b^2 v^2(\cos^2\chi+\sin^2\chi\cos^2\gamma)-a^2 v^2\cos^2\gamma.
\ee
The constants $E$ and $L_z$ are the energy and the $z$ component of the angular momentum of the particle, respectively, and $C$ is the Carter's constant. The critical impact parameter is found by imposing $R(r)=0$ and $R'(r)=0$. In Sec. \ref{sec:res} we compare $\sigma_{\text{hf}}$ with the numerical results for a selection of parameters, including the incidence angle.

\section{Low-frequency limit}\label{sec:low}

It is possible to find an analytical solution for the radial equation (\ref{eq:radialeq}) in the low-frequency limit. For the massless (and chargeless) scalar case, the absorption cross section goes to the area of the black hole in this limit, a result which was shown to be valid quite generically \cite{Higuchi:2001si}. We can also find an analytical solution for the massive and charged case, restricting our analysis to low values of the scalar field mass ($\mu\approx 0$) and charge ($q\approx 0$). For small frequencies, the greatest contribution comes from the mode for $l=m=0$, such that we will find an approximate solution only for this case.

In order to obtain the low-frequency limit we will first find a solution for the null frequency case and then match this solution with the asymptotic limits, given by Eq. (\ref{inmodes}). Considering $\omega=0$ (and also $\mu=q=l=m=0$) in Eq. (\ref{eq:radialeq}), we find the solution 
\b
U^{(0)}=\sqrt{r^2+a^2}\left\{A -\frac{B}{2} \left[\ln\frac{(r-r_+)}{(r-r_-)}+i\pi\right]\right\},
\label{eq0}
\e
where $A$ and $B$ are constants to be determined. For the limit $r\rightarrow r_+$, Eq. (\ref{eq0}) reduces to 
\b
U^{(0)}\approx\sqrt{r_+^2+a^2}\left\{A -\frac{B}{2} \left[\ln\frac{(r-r_+)}{(r_+-r_-)}+i\pi\right]\right\}.
\label{eq0h}
\e

Close to the horizon we can consider the solution given by Eq. (\ref{inmodes}), keeping only the term $j=0$ of the summation in Eq. (\ref{eq:trw}), such that we have
\b
U^H={\mathcal{T}_{\omega 00}} e^{-i(\omega-\omega_c)r_\star}.
\label{sho}
\e
In this limit the tortoise coordinate behaves as
\b
r_\star \approx  \alpha\ln(r-r_+),
\label{toh}
\e
where 
\b
\alpha \equiv  \frac{(r_+^2+a^2)}{2\sqrt{M^2-(a^2+Q^2)}}.
\e
Considering Eq. (\ref{sho}), together with Eq. (\ref{toh}), in the limit $\omega \rightarrow 0$, we have
\b
U^H \approx {\mathcal{T}_{\omega 00}} [1-i(\omega-\omega_c)\alpha \ln(r-r_+)].
\label{eqh0}
\e

Comparing Eqs. (\ref{eq0h}) and (\ref{eqh0}) we find
\bea
A&=&\frac{{\mathcal{T}_{\omega 00}}}{\sqrt{r_+^2+a^2} }\left[1-\pi(\omega-\omega_c)\alpha\right],\label{coea}\\
B&=&\frac{2i{\mathcal{T}_{\omega 00}}(\omega-\omega_c)\alpha }{\sqrt{r_+^2+a^2}},
\label{coe}
\eea
where we neglected a term of $(\omega-\omega_c) \ln(r_+-r_-)$ to obtain Eq. (\ref{coea}).

In the limit $r\rightarrow \infty$, the radial equation (\ref{eq:radialeq}) is given by
\be
\l(\frac{d^2}{d r^2}+V^{\infty}\r)U^\infty=0,
\label{eq:rei}
\ee
where
\b
V^{\infty}=(\omega^2-\mu^2)+\frac{2(2M\omega^2-M\mu^2-Qq\omega)}{r}.
\label{poti}
\e
To obtain Eqs. (\ref{eq:rei}) and (\ref{poti}) we neglected terms of order $\mathcal{O}(r^{-2})$.
The solution for Eq. (\ref{eq:rei}) can be written as
\b
U^\infty = \zeta F_0(\eta, \omega v r)+ \kappa G_0(\eta,\omega v r),
\label{eqi}
\e
where $F_l(\eta,x)$ and $G_l(\eta,x)$ are the Coulomb wave functions and
\b
\eta \equiv -\frac{M(2\omega^2-\mu^2)}{\omega v}+\frac{Qq}{v} = -\frac{M\mu(1+v^2)}{v\sqrt{1-v^2}}+\frac{Qq}{v}.
\label{eta}
\e
The coefficients $\zeta$ and $\kappa$ of Eq. (\ref{coe}) are related to ${\mathcal{I}_{\omega lm}}$ and ${\mathcal{R}_{\omega lm}}$ through 
\b
{\mathcal{I}_{\omega 0 0}}=\frac{-\zeta+i\kappa}{2i}
\label{inc}
\e
and
\b
{\mathcal{R}_{\omega 0 0}}=\frac{\zeta+i\kappa}{2i}.
\label{ref}
\e

Considering the limit $\omega v r \ll 1$ in Eq. (\ref{eqi}), we find
\b
U^\infty \approx \zeta \beta \omega v r +\frac{\kappa}{\beta},
\label{eqi0}
\e
where we used the approximations $F_0(\eta,x)=\beta x$ and $G_0(\eta,x)=1/\beta$, with
\b
\beta^2 \equiv \frac{2\pi \eta}{e^{2\pi \eta}-1}.
\label{beta}
\e

Considering the limit $r \gg r_+$ in Eq. (\ref{eq0}), we find
\b
U^0 \approx \left(A-\frac{iB\pi}{2} \right)r+B\sqrt{M^2-(a^2+Q^2)}.
\label{eq0i}
\e
Comparing Eqs. (\ref{eqi0}) and (\ref{eq0i}), and using also Eq. (\ref{coe}), we find
\bea
\zeta&=& \frac{{\mathcal{T}_{\omega 00}}}{\beta \omega v \sqrt{r_+^2+a^2}}, \nonumber\\
\kappa&=& i{\mathcal{T}_{\omega 00}}(\omega-\omega_c)\beta\sqrt{r_+^2+a^2}.
\label{aeb}
\eea

Substituting Eqs. (\ref{aeb}) in Eqs. (\ref{inc}) and (\ref{ref}), we obtain
\b
{\mathcal{I}_{\omega 00}}=-\frac{{\mathcal{T}_{\omega 00}}[1+\beta^2(r_+^2+a^2) \omega (\omega-\omega_c)v]}{2i \beta \omega v \sqrt{r_+^2+a^2}},
\label{int}
\e

\b
{\mathcal{R}_{\omega 00}}=\frac{{\mathcal{T}_{\omega 00}}[1-\beta^2(r_+^2+a^2) \omega (\omega-\omega_c)v]}{2i \beta \omega v \sqrt{r_+^2+a^2}}.
\label{ret}
\e

Substituting Eqs. (\ref{int}) and (\ref{ret}) in Eq. (\ref{abs}), and using that, in the low-frequency limit, $S_{\omega 0 0}\rightarrow 1/\sqrt{4\pi}$, we find
\b
\sigma\approx \frac{\pi}{\omega^2 v^2}\left\{\frac{4 \beta^2(r_+^2+a^2) \omega (\omega-\omega_c)v}{[1+\beta^2(r_+^2+a^2) \omega (\omega-\omega_c)v]^2}\right\}.
\e
By considering only the first term in the denominator, we are left with
\b
\sigma \approx \frac{4\pi (r_+^2+a^2) (\mu-\omega_c) \beta^2}{\mu v}.
\e

Next, we have to consider the expansion of $\beta$, which, according to Eqs. (\ref{eta}) and (\ref{beta}), depends on the velocity of the scalar field, and we define $v_c\equiv 2\pi(M\mu-Qq)$. 

For $v>v_c$, $\beta \rightarrow 1$, we obtain
\be
\sigma^{(1)}_{\text{lf}}=\frac{\mathcal{A}}{\mu v}\left(\mu-\frac{Qq r_+}{r_+^2+a^2}\right).
\label{lf1}
\ee
This result reduces to the one obtained in Ref. \cite{Benone:2015bst} for $a\rightarrow 0$ and to the result of Ref. \cite{Benone:2014qaa} for both $a$ and $q$ going to zero. 

For $v<v_c$, $\beta \rightarrow 2\pi(M\mu-Qq)/v$, we obtain
\be
\sigma^{(2)}_{\text{lf}}=\frac{8\left(\pi\sqrt{r_+^2+a^2}\right)^2}{\mu v^2}\left(\mu-\frac{Qq r_+}{r_+^2+a^2}\right)\left(M\mu-qQ\right).
\label{lf2}
\ee
This result reduces to the one in Ref. \cite{PhysRevD.95.044035} for $a=q=0$.



\begin{figure}%
\includegraphics[width=\columnwidth]{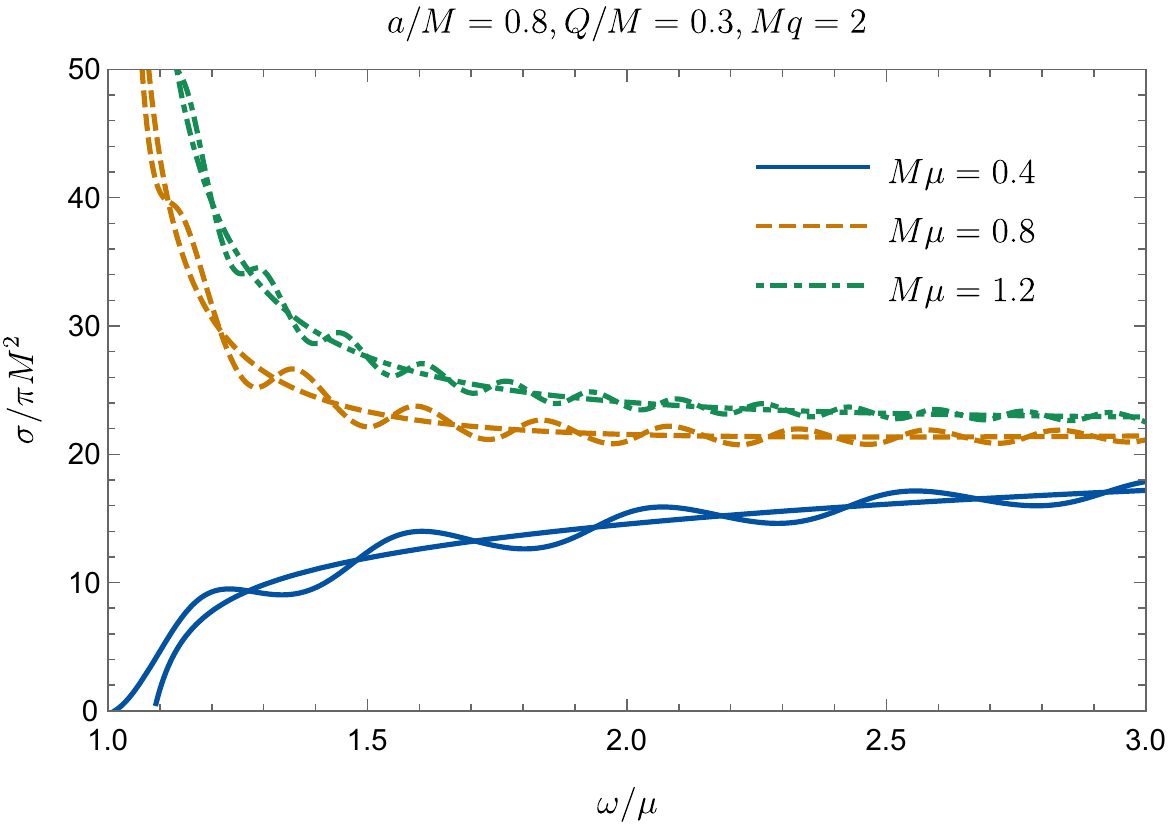}\\
\includegraphics[width=\columnwidth]{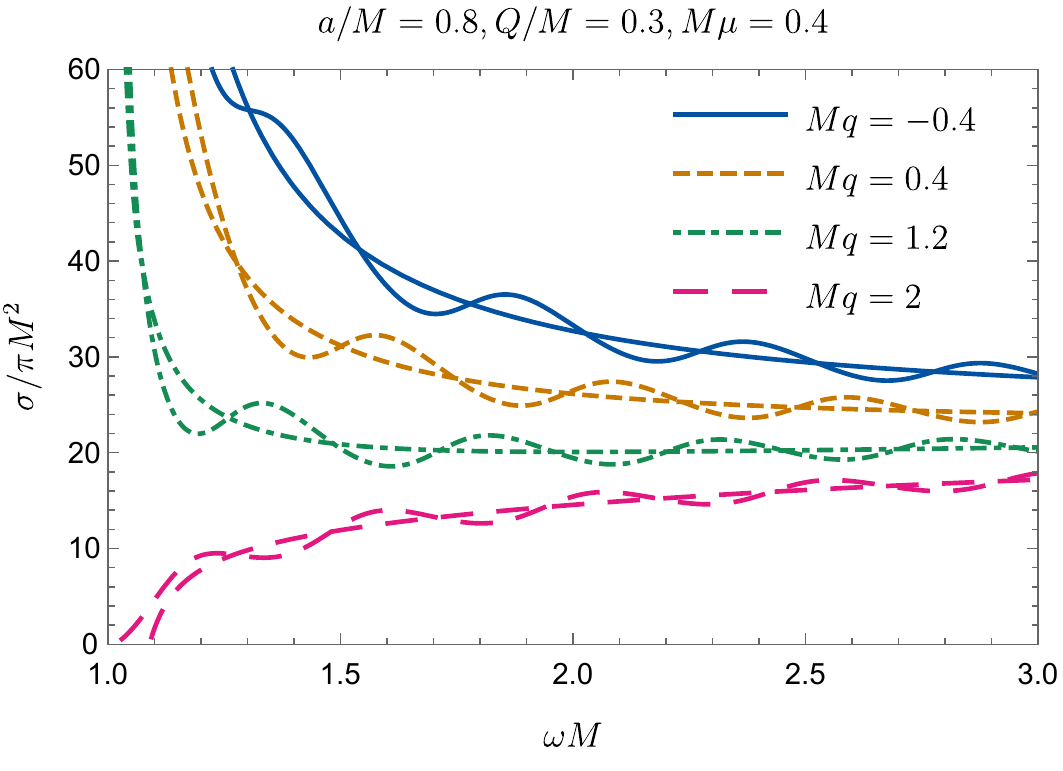}
\caption{Total absorption cross section for on-axis incidence, for different values of the scalar field mass (top) and charge (bottom). The numerical solutions oscillate around the high-frequency limit.}%
\label{tot}%
\end{figure}

\section{Results}\label{sec:res}

In this section we present an assortment of our numerical results, which we compare with the high- and low-frequency limits obtained in previous sections. We solved Eq. (\ref{eq:radialeq}) numerically, using the asymptotic expressions given by Eqs. (\ref{inmodes}), with the help of Eqs. (\ref{eq:inc}) and (\ref{eq:trw}). In order to simplify the calculations we assume $h_0=g_0=1$, while the other terms of $h_j$ and $g_j$ in Eqs. (\ref{eq:inc}) and (\ref{eq:trw}), respectively, are found by assuming that the radial equation is subjected to the boundary conditions given by Eq. (\ref{inmodes}). Knowing the radial solution, we can find the reflection coefficient and, hence, the absorption cross section through Eq. (\ref{eq:abs}).

In Fig. \ref{tot} we present the total absorption cross section for incidence along the black hole rotation axis, varying the parameters of the scalar field ($\mu$ and $q$). The numerical results oscillate around the corresponding high frequency solution, tending to it as we increase $\omega$. As we increase the mass of the field, the absorption cross section increases. However, as we increase the charge of the field, the absorption cross section decreases. For some ranges of the parameters, the total absorption cross section goes to infinity as $\omega \rightarrow \mu$, but as we increase the value of $q$ we obtain finite values of the absorption cross section for this limit. 

\begin{figure}%
\includegraphics[width=\columnwidth]{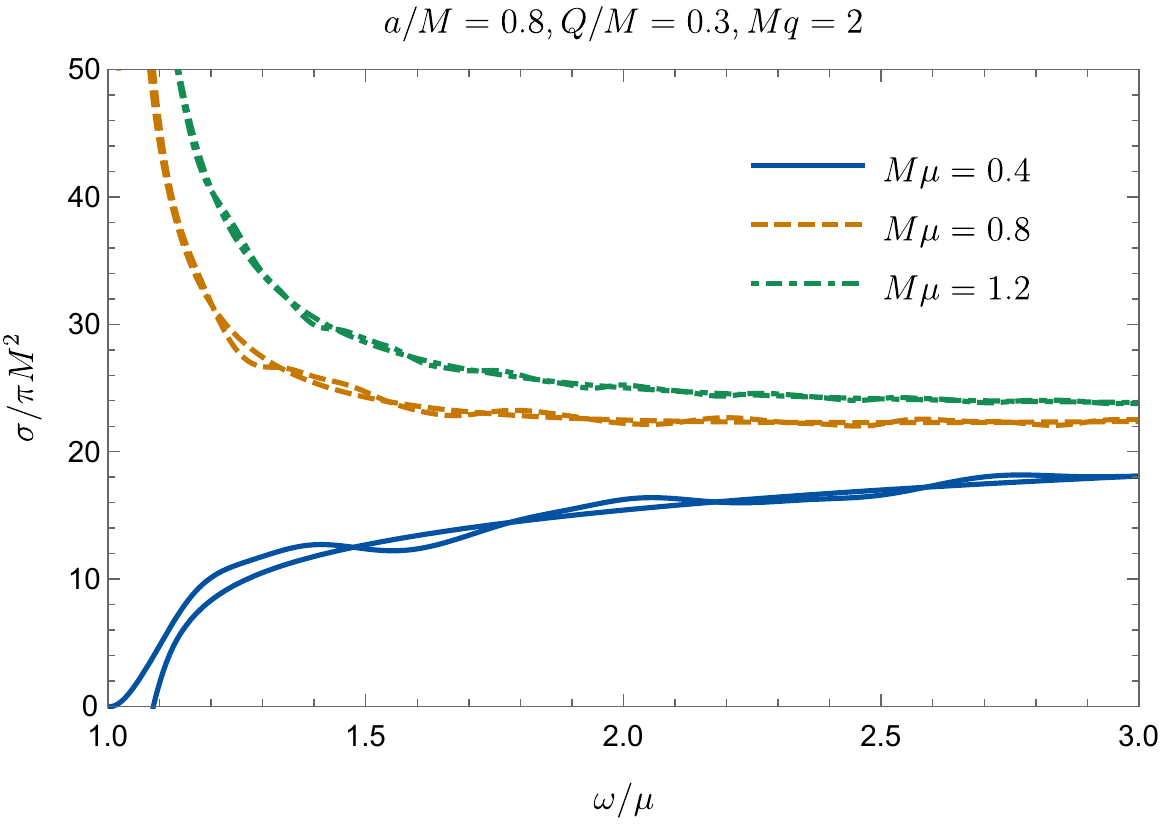}\\
\includegraphics[width=\columnwidth]{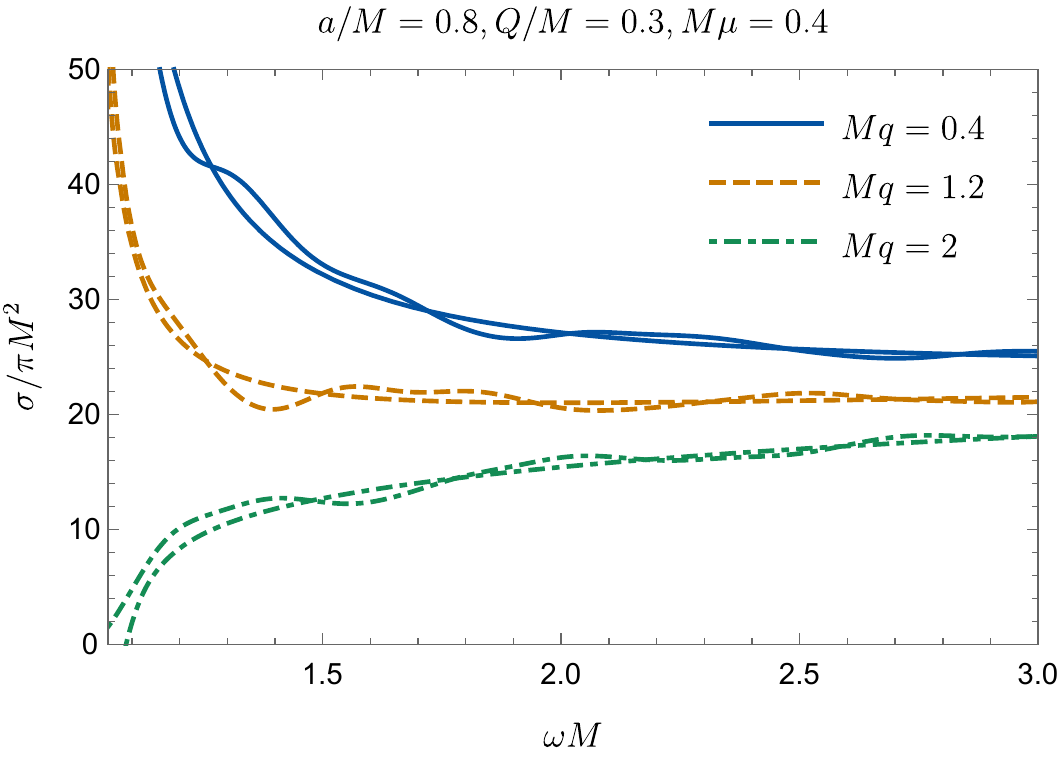}
\caption{Total absorption cross section for equatorial incidence ($\gamma=\pi/2$), for different values of the scalar field mass (top) and charge (bottom). The numerical solutions oscillate around the high-frequency limit in a less regular fashion than in Fig. \ref{tot}.}%
\label{tot90}%
\end{figure}

\begin{figure}%
\includegraphics[width=\columnwidth]{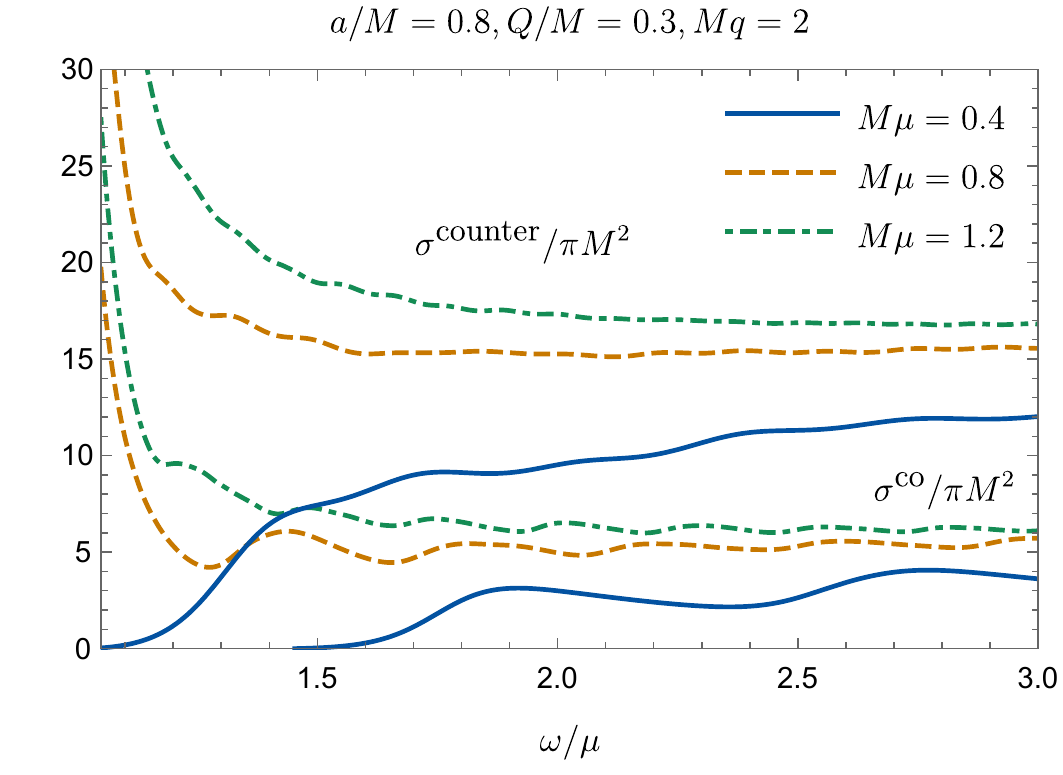}\\
\includegraphics[width=\columnwidth]{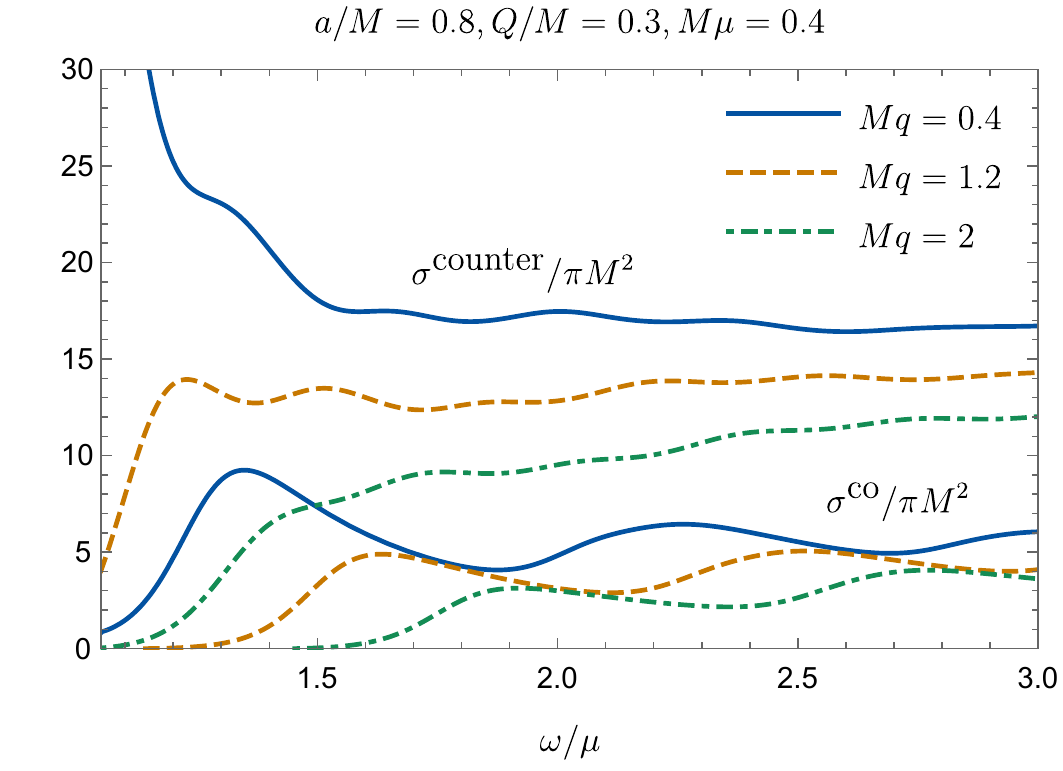}
\caption{Corotating [Eq. (\ref{coro})] and counterrotating [\ref{counter}] contributions to the absorption cross section, for equatorial incidence ($\gamma=\pi/2$) and different values of the scalar field mass (top) and charge (bottom).}%
\label{cc}%
\end{figure}

For the off-axis incidence case, the oscillations of the total absorption cross section become less regular as we increase $\gamma$, specially for high values of $a$. In Fig. \ref{tot90} we present results for the off-axis incidence in the equatorial plane ($\gamma=\pi/2$). The behavior of the off-axis incidence can be better seen if we consider the corotating [given by Eq. (\ref{coro})] and the counterrotating [given by Eq. (\ref{counter})] contributions separately, as shown in Fig. \ref{cc}. The counterrotating modes are always more absorbed than the corotating ones. Depending on the values of the parameters, both modes can go to infinity or to a finite value in the low-frequency limit.

\begin{figure}%
\includegraphics[width=\columnwidth]{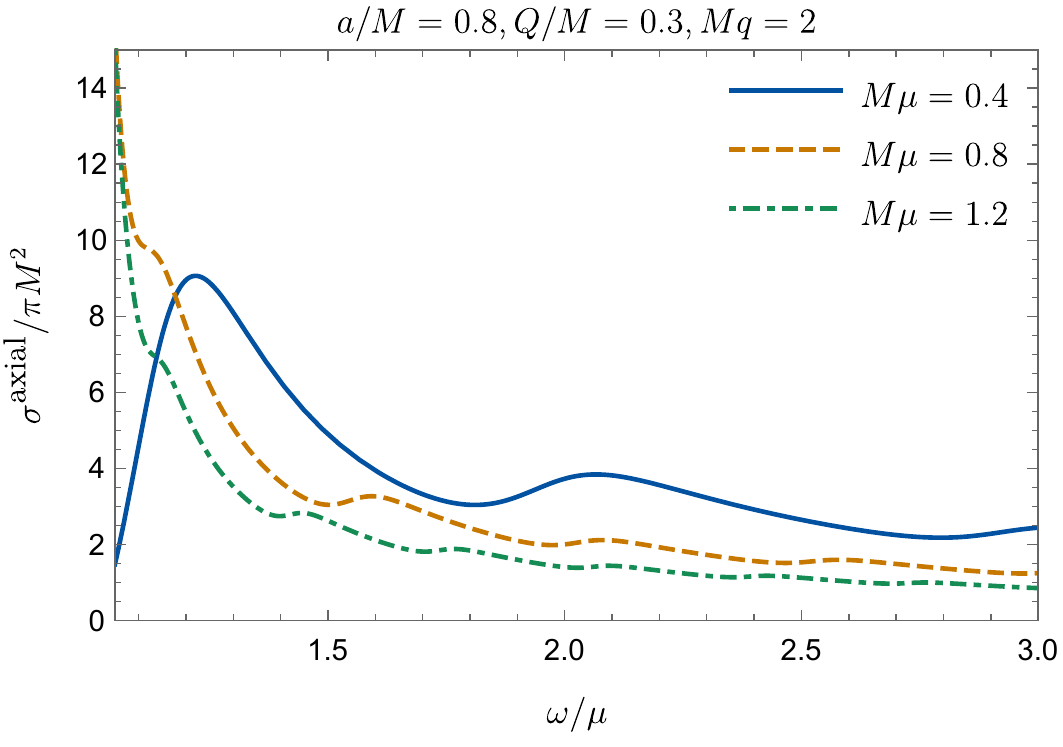}\\
\includegraphics[width=\columnwidth]{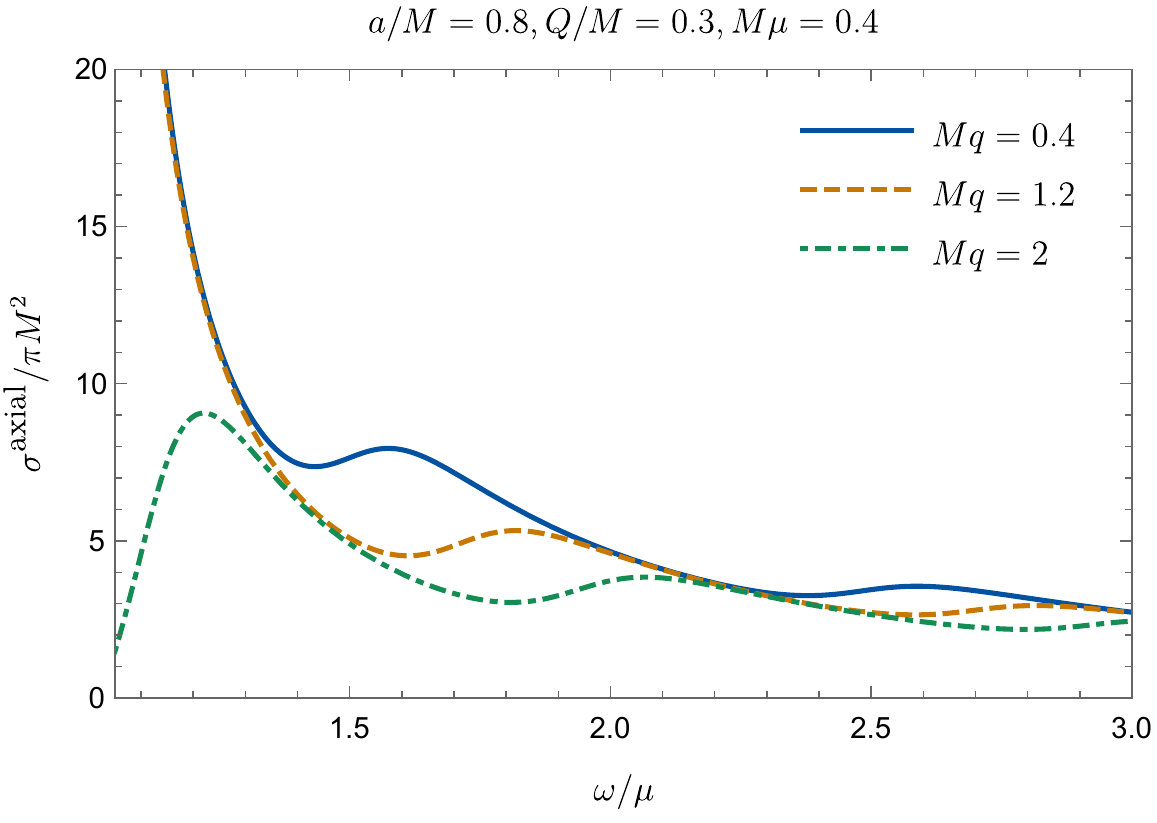}
\caption{Axial contribution [Eq. (\ref{axial})] to the absorption cross section, for equatorial incidence ($\gamma=\pi/2$) and different values of the scalar field mass (top) and charge (bottom).}%
\label{ax}%
\end{figure}

In Fig. \ref{ax} we display the contribution to the absorption cross section given by the axiallly-symmetric modes [cf. Eq. (\ref{axial})]. In general, as we increase either the charge or the mass of the field, the axial absorption cross section increases. However, in the low-frequency limit, as we decrease the mass of the field, the axial absorption cross section increases.

\begin{figure}%
\includegraphics[width=\columnwidth]{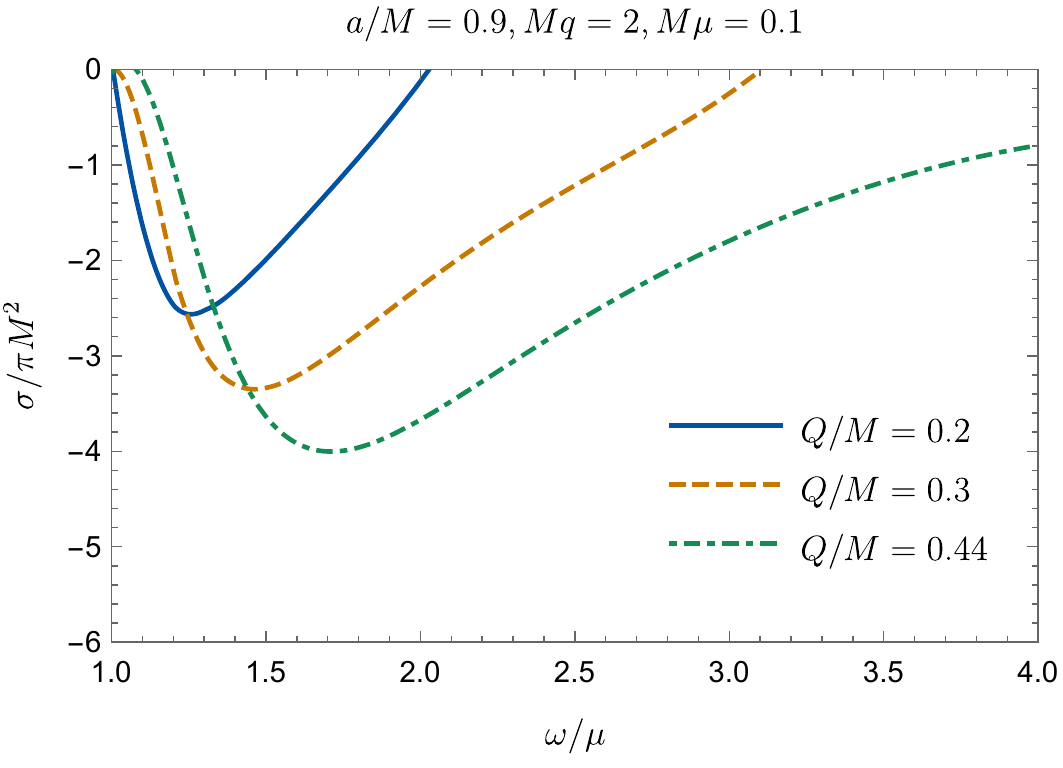}
\caption{Negative part of the total absorption cross section, for equatorial incidence ($\gamma=\pi/2$) and different values of the black hole charge.}%
\label{sup}%
\end{figure}

\begin{figure}%
\includegraphics[width=\columnwidth]{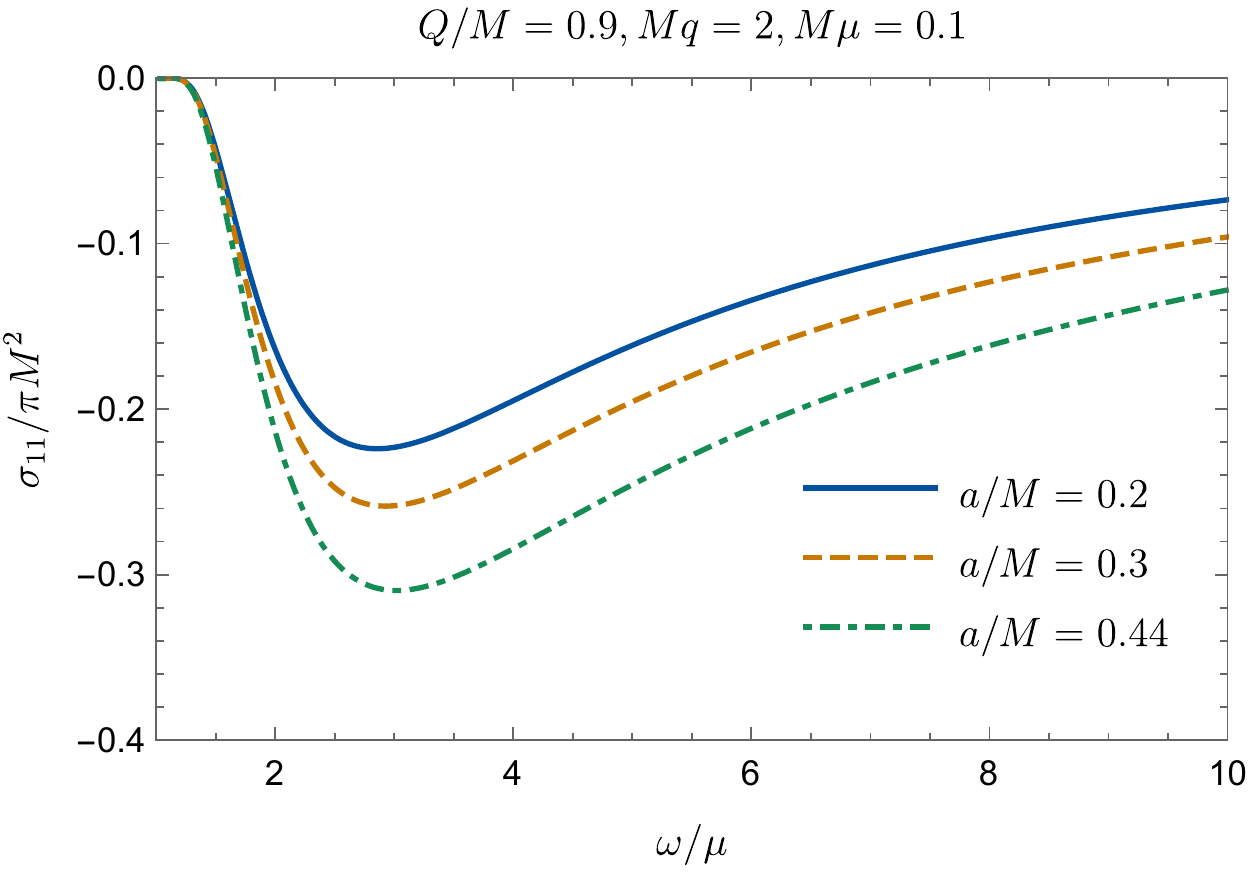}
\caption{Negative part of the partial absorption cross section, for $l=m=1$, $\gamma=\pi/2$ and different values of the black hole angular momentum.}%
\label{supa}%
\end{figure}

In Fig. \ref{sup} we exhibit only the negative part of the total absorption cross section, for different values of the black hole charge. We see that, as we increase $Q$, the superradiance increases. In Fig. \ref{supa} we also exhibit the negative part of the partial absorption cross section, this time for $l=m=1$, and for different values of the black hole angular momentum. As we increase $a$, the superradiance increases, but the increment with $Q$ is more significant [cf. Fig. \ref{sup}]. 

\begin{figure}[H]%
\includegraphics[width=\columnwidth]{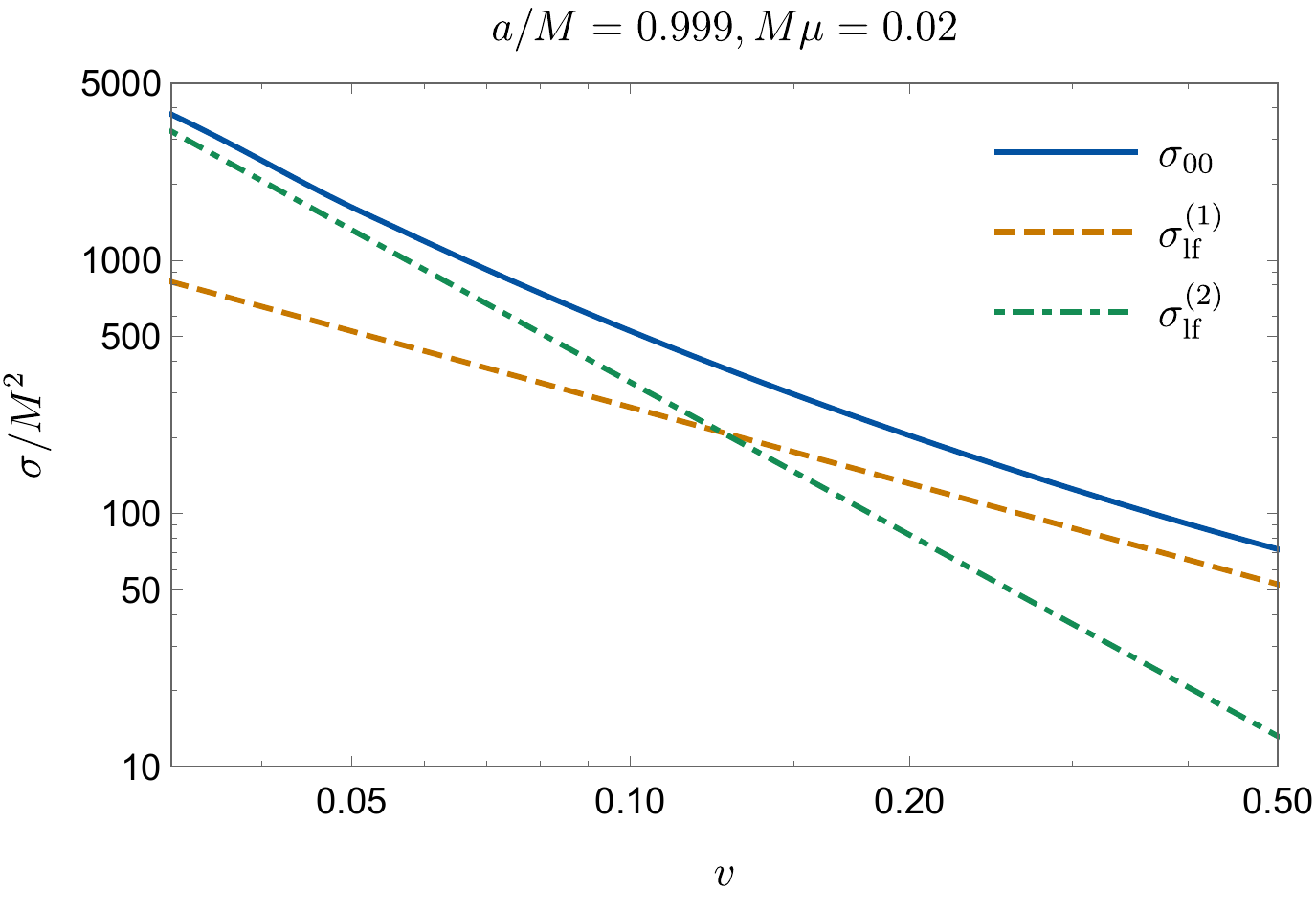}\\
\includegraphics[width=\columnwidth]{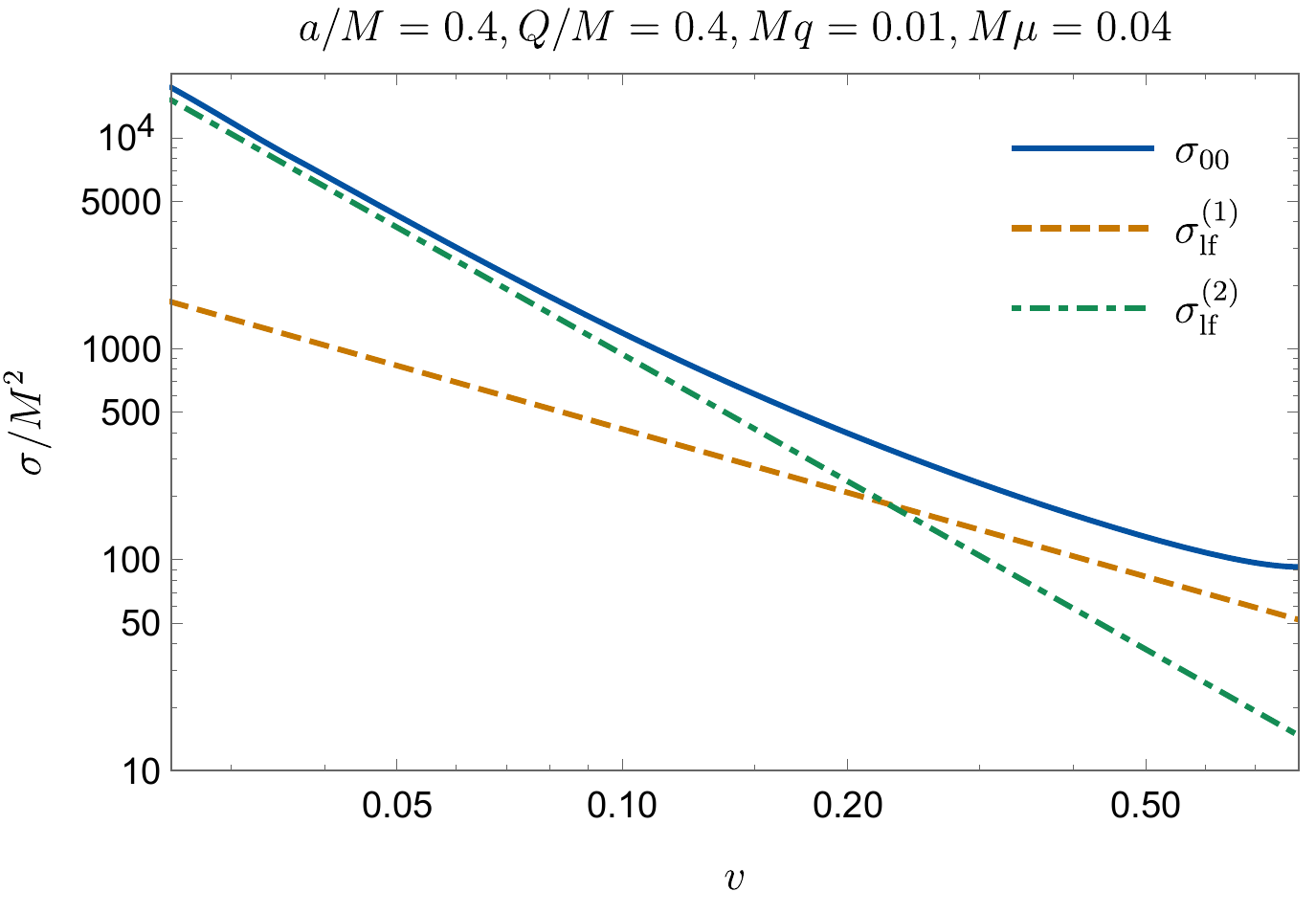}
\caption{Partial absorption cross section for $l=m=0$, compared with their low-frequency limit analytical approximations, given by Eqs. (\ref{lf1}) and (\ref{lf2}), for the Kerr (top) and Ker-Newman (bottom) spacetimes. For the top plot we have $v_c=0.13$, while for the bottom plot $v_c=0.23$.}%
\label{lfa}%
\end{figure}

In Fig. \ref{lfa} we display the low-frequency limits of the absorption cross section, given by Eqs. (\ref{lf1}) and (\ref{lf2}), comparing these with numerical results for the partial absorption cross section for $l=m=0$. In the top plot of Fig. \ref{lfa} we consider the chargeless BH case, while in the bottom we consider the charged BH case. For both cases we see a transition of the numerical results for the cross section from the $\sigma_{\text{lf}}^{(1)}$ to the $\sigma_{\text{lf}}^{(2)}$ limits at $v\approx v_c$.

\section{Final remarks}\label{sec:remarks}

We have considered a massive and charged scalar field around a Kerr-Newman black hole, computing the partial and total absorption cross sections, using the partial wave method. The limits for low frequencies were obtained analytically, while the high-frequency limit was obtained semi-analytically. We obtained numerically the total absorption cross section for the full range of frequencies, comparing our numerical results with the (semi-)analytical limits in the suitable regimes, finding excellent agreement.

For the on-axis incidence case we have shown that the total absorption cross section oscillates around the high-frequency limit, a feature already known from other black hole cases such as for static spacetimes. For $Qq>0$, as we increase $\mu$ and decrease $q$, the total absorption cross section decreases. This happens because the term $Qq$ implies the presence of a repulsive electromagnetic force, which decreases the absorption. When $Qq<0$, an electromagnetic attractive force is present, increasing the absorption. For a massive scalar field the total absorption cross section goes to infinity as $\omega\rightarrow \mu$; however, for $Qq>0$ the total absorption cross section can have a finite low-frequency limit, depending on the values of the parameters.

For off-axis incidence we have shown that the total absorption cross section presents less regular oscillations around the high-frequency limit, a feature which can be better understood when we consider the corotating and counterrotating contributions separately. We obtained that the counterrotating modes are more absorbed than the corotating ones, due to the dragging of inertial frames in the rotating black hole spacetime.

We also investigated superradiance, showing that, as we increase either $a$ or $Q$, the superradiance increases. For the chargeless black hole case the total absorption cross section is always positive, while for the charged black hole case the total absorption cross section can be negative. This becomes explicit when we consider the contributions of corotating and counterrotating modes separately. For the chargeless case the absorption cross section associated to the counterrotating modes is always positive, what accounts, when summed with the corotating modes, for a positive total absorption cross section. For the charged black hole case both corotating and counterrotating modes can present a negative absorption cross section, due to the influence of the Lorentz force, resulting in a negative total absorption cross section for low frequencies. 

For the low-frequency limit we have shown that the absorption cross section has two different limits, depending on the velocity of the field, as already shown for other cases \cite{Unruh:1976fm, PhysRevD.95.044035}. We have compared the numerical and analytical results, showing that the numerical absorption cross section presents a transition of regime near a critical velocity, $v_c$. Since for massive fields the low-frequency limit corresponds to $\omega \rightarrow \mu$, then this regime can only be achieved by bosons with very low mass. It is possible to show that both regimes can be relevant in the context of ultralight bosons \cite{Hui:2016ltb}.



\begin{acknowledgments}

This study was financed in part by Coordenação de Aperfeiçoamento de Pessoal de Nível Superior (CAPES, Brazil) – Finance Code 001, and by Conselho Nacional de Desenvolvimento Científico e Tecnológico (CNPq, Brazil). 
This research has also received funding from the European Union's Horizon 2020 research and innovation programme under the H2020-MSCA-RISE-2017 Grant No. FunFiCO-777740.
The authors also acknowledge Luiz Carlos dos Santos Leite and Sam Dolan for useful discussions.
 
\end{acknowledgments}

\bibliography{refs}

\end{document}